\documentclass[journal]{IEEEtran}
\usepackage{amsmath, amssymb, amsthm}

\usepackage{tikz}
\usetikzlibrary{positioning}
\usetikzlibrary{shapes.geometric, arrows.meta, positioning}
\usetikzlibrary{shapes}
\usepackage{comment}
\usepackage{graphicx} 
\usepackage{subcaption}
\usepackage{hyperref}
\usepackage{booktabs, tabularx, siunitx}
\newtheorem{remark}{Remark}

\usepackage[font=scriptsize]{caption}
\usepackage[normalem]{ulem}
\usepackage{bm}

\usepackage[none]{hyphenat} 
\usepackage{microtype}      
\begin{document}
\title{Learning Constraint Surrogate Model for Two-stage Stochastic Unit Commitment}

\author{Amir~Bahador~Javadi,~\IEEEmembership{Graduate~Student~Member,~IEEE,}
        Amin~Kargarian,~\IEEEmembership{Senior~Member, IEEE, and}
        Mort Naraghi-Pour,~\IEEEmembership{Life~Senior~Member, IEEE}
        \vspace{-15pt}

\thanks{This material is based upon work supported by the National Science Foundation under Grant Number 1944752.}
\thanks{The authors are with the Department of Electrical and Computer Engineering, Louisiana State University, Baton Rouge,
LA, 70802, USA e-mail: \url{ajavad2@lsu.edu}, \url{kargarian@lsu.edu}, \url{naraghi@lsu.edu}.}
}

\maketitle

\begin{abstract}
The increasing penetration of renewable energy sources introduces significant uncertainty in power system operations, making traditional deterministic unit commitment approaches computationally expensive. This paper presents a machine learning surrogate modeling approach designed to reformulate the feasible design space of the two-stage stochastic unit commitment (TSUC) problem, reducing its computational complexity. The proposed method uses a support vector machine (SVM) to construct a surrogate model based on the governing equations of the learner. This model replaces the original $2|\mathcal{L}| \times |\mathcal{S}|$ transmission line flow constraints—where $|\mathcal{S}|$ is the number of uncertainty scenarios and $|\mathcal{L}|$ is the number of transmission lines, with $|\mathcal{S}| \ll |\mathcal{L}|$—with a significantly reduced set of $1 \times |\mathcal{S}|$ linear inequality constraints. The approach is theoretically grounded in the polyhedral structure of the feasible region under DC power flow approximation, enabling the transformation of $2|\mathcal{L}|$ line flow limit constraints into a single linear constraint. The surrogate model is trained using data generated from computationally efficient DC optimal power flow simulations. Simulation results on the IEEE 57-bus and 118-bus systems demonstrate SVM halfspace constraint accuracy of 99.72\% and 99.88\%, respectively, with TSUC computational time reductions of 46\% and 31\% and negligible generation cost relative  (0.63\%, and 0.88\% on average for IEEE 57- and 118-bus systems, respectively). This shows the effectiveness of the proposed approach for practical power system operations under renewable energy uncertainty.
\end{abstract}

\begin{IEEEkeywords}
Two-stage stochastic unit commitment, machine learning, support vector machines, surrogate modeling.
\end{IEEEkeywords}

\IEEEpeerreviewmaketitle

\section{Introduction}
\IEEEPARstart{P}{ower} system optimization presents a multifaceted challenge encompassing various computational problems that operate across different temporal scales. These problems include optimal power flow, economic dispatch, unit commitment (UC), and many other optimization tasks that must be solved hourly, daily, weekly, or over extended planning horizons \cite{wood2013power}. Among these challenges, the UC problem is particularly demanding due to its combinatorial nature and the need for high-resolution daily scheduling cycles to determine the optimal scheduling of generating units \cite{padhy2004unit}.

The complexity of UC extends beyond its inherent computational demands, as UC formulations frequently emerge as subproblems within broader optimization frameworks that are not explicitly UC problems \cite{conejo2006decomposition}. In addition to that, the increasing penetration of renewable energy resources, particularly wind and solar, introduces significant uncertainty in both generation and load forecasts, substantially amplifying the computational burden of traditional UC formulations. These challenges have made it essential to develop methods that can efficiently and accurately address uncertainty in evolving power systems \cite{vanackooij2018large, energies_renewable_uc_2017}.

Several approaches have been developed to address uncertainty in UC problems, including robust optimization \cite{bertsimas2013adaptive}, chance-constrained programming \cite{lubin2016chance}, and stochastic programming techniques \cite{takriti1996stochastic}. Two-stage stochastic unit commitment (TSUC) has emerged as a promising solution for addressing uncertainty due to its structural compatibility with the sequential decision-making nature of power systems \cite{shiina2004stochastic}.

The TSUC framework embodies the \textit{here-and-now} and \textit{wait-and-see} decision paradigm, which closely mirrors the operational reality of power system dispatch \cite{birge2011introduction}. In this framework, first-stage decisions (such as unit commitment schedules) are made before uncertainty is resolved, while second-stage decisions (such as economic dispatch) are made after uncertainty realization. This temporal structure provides a more realistic representation of power system operations compared to deterministic approaches \cite{morales2013integrating}.

Despite its theoretical advantages, the computational intensity of TSUC remains a significant barrier to practical implementation, often requiring the solution of large-scale mixed-integer programs over a set of uncertainty scenarios \cite{tahanan2015large}. Various computational enhancement strategies have been proposed, including distributed unit commitment algorithms \cite{li2013distributed} and decomposition techniques \cite{guan2003lagrangian}. While these methods can achieve marginal or substantial computational improvements, the fundamental need for further complexity reduction persists 
to enable deployment in increasingly complex and realistic power system models \cite{vanackooij2018large}.

Recent advances in machine learning have opened new avenues for addressing computational challenges in power system optimization. Researchers have explored learning-based approaches for predicting UC outcomes, reducing problem dimensionality, or approximating constraint sets \cite{zamzam2020learning, 9282137}. However, many of these machine learning approaches suffer from the \textit{black box} limitation, where the lack of interpretability and integration with physics-based constraints obscures the underlying physical relationships, raising concerns about their reliability in safety-critical power systems and potentially undermining operator confidence \cite{donnot2020interpretable}.

Alternative approaches that preserve physical interpretability include inactive constraint identification \cite{bienstock2014using}, machine learning-based optimization initialization \cite{li2018machine}, model reduction techniques \cite{wang2020model}, and explicit data-driven surrogate constraints for small-signal stability \cite{9652094}. These approaches come before solving the optimization problem as a separate pre-process in problem formulation and aim to improve computational efficiency while maintaining some degree of physical transparency. Nonetheless, they often lack strong theoretical guarantees and are typically limited to specific problem structures. Also, these learning strategies do not aim to replace computationally expensive constraints with easier-to-solve ones. 

Despite these efforts, integrating ML into power grid optimization in a physically grounded, computationally efficient manner remains an open challenge. Fully replacing an entire optimization model with a learner is often neither practical nor efficient. Instead, a more promising direction involves selectively modeling only the computationally intensive components of an optimization problem. For example, network constraints increase the complexity of the TSUC problem. Developing surrogate models to represent these constraints—while ensuring compatibility with the overall optimization framework—can reduce problem size and computational burden. However, embedding such ML-based surrogate models into the original optimization seamlessly and theoretically soundly remains a critical challenge.

To address this gap, this paper presents an approach grounded within the framework of symbolic regression, representing a middle ground between implicit machine learning methods and traditional applied mathematics \cite{udrescu2020ai}. Symbolic regression enables the data-driven discovery of explicit mathematical expressions that describe system dynamics or relationships \cite{JAVADI2025116075}. Our approach focuses on developing surrogate models for computationally intensive components within the optimization framework, seeking to derive governing equations that can be seamlessly integrated into the optimization problem \cite{brunton2016discovering}.

Specifically, we propose a support vector machine (SVM) based approach that transforms the feasibility determination problem into an explicit halfspace representation. This halfspace, defined by power generation and wind generation variables, provides a closed-form expression that can be directly incorporated into the optimization framework \cite{pourahmadi2025unit}. The resulting linear decision boundary enables rapid feasibility assessment while maintaining mathematical rigor and physical interpretability, thereby bridging the gap between machine learning efficiency and optimization transparency. The SVM formulation allows for efficient dual optimization while preserving the convex structure necessary for integration with power system optimization problems \cite{bennett2006dual}.

The main contributions of this work are threefold. First, we propose a surrogate modeling approach that replaces $2|\mathcal{L}| \times |\mathcal{S}| \times |\mathcal{T}| $ traditional line flow limit constraints with $1 \times |\mathcal{S}| \times |\mathcal{T}| $ inequality constraints derived from SVM hyperplane representation, where $|\mathcal{S}|$ is the number of uncertainty scenarios, $|\mathcal{L}|$ is the number of transmission lines, and $|\mathcal{T}|$ is the number of hours. This consolidation significantly reduces the constraint dimensionality while maintaining feasibility guarantees for transmission line limits. Second, we demonstrate the seamless integration of this surrogate model into the TSUC framework, where the hyperplane-based feasibility assessment can be efficiently incorporated into both first-stage commitment decisions and second-stage dispatch operations. Third, we propose a data generation method that circumvents the computational burden of solving full unit commitment problems by utilizing only DCOPF solutions for training the surrogate model. This approach significantly reduces the computational overhead associated with generating the training dataset while preserving the essential characteristics of the feasibility region. These contributions enable a computationally cheaper solution for TSUC problems under renewable energy and load uncertainty.

The remainder of this paper is organized as follows. Section~\ref{s2} formulates the TSUC problem under wind power and load uncertainty. Section~\ref{s3} introduces the proposed surrogate modeling framework, providing theoretical justification for replacing line flow constraints with a halfspace feasibility approximation. Section~\ref{2} describes the data generation process and simulation results. Section~\ref{s4} discusses the key findings and practical implications. Section~\ref{s5} provides concluding remarks.

\section{Two-Stage Stochastic Unit Commitment (TSUC)}
\label{s2}
The UC problem determines the optimal ON/OFF scheduling and generation levels for thermal generators over a finite planning horizon (typically 24 hours), minimizing total operational cost while satisfying demand and technical constraints. Traditional deterministic UC formulations may lead to infeasibility or economic inefficiency under uncertain renewable generation or load forecasts \cite{padhy2004unit, takriti1996stochastic}. To address these challenges, a two-stage stochastic programming approach has been widely adopted in the literature \cite{morales2013integrating}. 

\subsection{Notation}

\textbf{Sets:} Let $\mathcal{G}$ denote the set of all generators in the system, and $\mathcal{G}_i$ represent the subset of generators located at bus $i$. The time horizon is divided into discrete intervals defined by the set $\mathcal{T}$, and $\mathcal{S}$ denotes the set of all uncertainty scenarios considered. The set of buses is denoted by $\mathcal{B}$, while $\mathcal{L}$ represents the set of transmission lines in the network. For each bus $i \in \mathcal{B}$, $\mathcal{N}_i$ denotes the set of buses directly connected to it via transmission lines.

\textbf{Parameters:} Each generator $g \in \mathcal{G}$ has minimum and maximum generation capacities denoted by $P_g^{\min}$ and $P_g^{\max}$, respectively. The ramp-up and ramp-down limits are represented by $\text{RU}_g$ and $\text{RD}_g$. Parameters $m_g^{\uparrow}$ and $m_g^{\downarrow}$ define the minimum up and down time requirements. The startup and shutdown costs of generator $g$ are given by $SU_g$ and $SD_g$, while $C_{g0}$, $C_{g1}$, and $C_{g2}$ denote the cost coefficients in the quadratic cost function. For each transmission line $(i,j) \in \mathcal{L}$, $x_{ij}$ indicates the line reactance, and $\bar{f}_{ij}$ is its thermal capacity limit. The net demand at bus $i$ at time $t$ under scenario $s$ is denoted by $D_{i,s,t}$. $P^{wind}_{s,t}$ is the renewable power generation. The probability of scenario $s$ occurring is given by $\pi_s$.

\textbf{Variables:} The binary variable $u_{g,t}$ indicates whether generator $g$ is committed at time $t$. The power generation of generator $g$ at time $t$ in scenario $s$ is denoted by $p_{g,s,t}$. The startup and shutdown of generator $g$ at time $t$ are captured by the binary indicators $y_{g,t}$ and $z_{g,t}$, respectively. The voltage angle at bus $i$, and bus $j$ at time $t$ in scenario $s$ are represented by $\theta_{i,s,t}$, and $\theta_{j,s,t}$, respectively. $\theta_{\text{ref},s,t}$ denotes the reference angle for each scenario and time, which is fixed at zero.

\subsection{Framework Overview}

In the TSUC framework, decisions are categorized into two stages:

\textbf{First-stage here-and-now decisions:} are made before the realization of uncertainty and include:
\begin{itemize}
    \item Unit commitment variables $u_{g,t}\in\{0,1\}$ indicating whether generator $g$ is online at time $t$.
    \item Startup and shutdown variables $y_{g,t}\in\{0,1\}$ and $z_{g,t}\in\{0,1\}$ indicating whether generator $g$ is started of shutdown at time $t$, respectively.
\end{itemize}

\textbf{Second-stage wait-and-see decisions:} are made after uncertainty (e.g., wind output or load variations) is realized and include:
\begin{itemize}
    \item Scenario-dependent recourse variables $p_{g,s,t}$ representing power generation in each scenario $s \in \mathcal{S}$.
    \item Scenario-dependent variables $\theta_{i,s,t}$ representing bus voltage angles in each scenario $s \in \mathcal{S}$.
\end{itemize}

Wind power uncertainty is typically modeled using a finite scenario set $\mathcal{S}$, where each scenario $s$ is characterized by its probability $\pi_s$ and a stochastic realization following $\mu_s + z_s \cdot \sigma_s$, where $\mu$, $\sigma$, and $z_s$ represent the mean, the standard deviation of the wind forecast, and a parameter ranging from -4 to 4, respectively \cite{morales2013integrating}. The objective of the two-stage formulation is to minimize the expected total operational cost across all scenarios and time periods, considering both first-stage scheduling decisions and second-stage adjustments made after uncertainty is realized.

\subsection{Mathematical Formulation}

The TSUC problem is formulated as a mixed-integer linear program (MIP). 

\begin{align}
\min \quad & \sum_{t=1}^{\mathcal{T}} \sum_{g \in \mathcal{G}} \left[ y_{g,t} SU_g + z_{g,t}SD_g \right] \nonumber \\
&+ \sum_{s \in |\mathcal{S}|} \pi_s \sum_{t=1}^{\mathcal{T}} \sum_{g \in \mathcal{G}} \left[ C_{g_2} p_{g,s,t}^2 + C_{g_1} p_{g,s,t} + u_{g,t}C_{g_0} \right] \label{eq:objective}
\end{align}

s.t.

First-stage Constraints:
\begin{align}
&y_{g,t} \geq u_{g,t} - u_{g,t-1}, \quad \forall g, t \geq 2 \label{eq:startup}\\
&z_{g,t} \geq u_{g,t-1} - u_{g,t}, \quad \forall g, t \geq 2 \label{eq:shutdown}\\
&u_{g,\tau} \geq y_{g,t}, \quad \forall \tau \in [t, t + m_g^{\uparrow} - 1], \quad \forall g, t \label{eq:minup}\\
&u_{g,\tau} \leq 1 - z_{g,t}, \quad \forall \tau \in [t, t + m_g^{\downarrow} - 1], \quad \forall g, t \label{eq:mindown}\\
&y_{g,t},z_{g,t},u_{g,t} \in \{0,1\}, \quad \forall g,t
\end{align}

Second-stage Constraints:
\begin{align}
&P_g^{\min} u_{g,t} \leq p_{g,s,t} \leq P_g^{\max} u_{g,t}, \quad \forall g, s, t \label{eq:capacity}\\
&- \text{RD}_g \leq p_{g,s,t} - p_{g,s,t-1} \leq \text{RU}_g, \quad \forall g, s, t \geq 2 \label{eq:ramping}\\
&\sum_{g \in \mathcal{G}_i} p_{g,s,t} + P^{wind}_{s,t} + \sum_{j \in \mathcal{N}_i} \frac{\theta_{i,s,t} - \theta_{j,s,t}}{x_{ij}} = D_{i,s,t}, \quad \forall i, s, t \label{eq:balance}\\
&\left| \frac{\theta_{i,s,t} - \theta_{j,s,t}}{x_{ij}} \right| \leq \bar{f}_{ij}, \quad \forall s, t, (i,j) \in \mathcal{L} \label{eq:transmission}\\
&\theta_{\text{ref},s,t} = 0, \quad \forall s, t \label{eq:reference}
\end{align}

The objective function in \eqref{eq:objective} minimizes the first-stage startup and shutdown costs along with the expected generation cost in the second stage.
The first-stage scenario-independent constraints include \eqref{eq:startup}--\eqref{eq:mindown}. Constraints \eqref{eq:startup}--\eqref{eq:shutdown} model the logical relationship between generator startup and shutdown decisions. Constraint \eqref{eq:startup} ensures that the startup indicator ($y_{g,t}$) is activated when a generator transitions from off to on state, while constraint \eqref{eq:shutdown} ensures that the shutdown indicator ($z_{g,t}$) is activated when a generator transitions from on to off state. Constraints \eqref{eq:minup}--\eqref{eq:mindown} enforce operational requirements that once a generator is started up, it must remain online for at least $m_g^{\uparrow}$ periods, and once it is shut down, it must remain offline for at least $m_g^{\downarrow}$ periods. Constraint \eqref{eq:minup} ensures minimum up-time compliance, while constraint \eqref{eq:mindown} ensures minimum down-time compliance.

The second-stage scenario-dependent constraints are \eqref{eq:capacity}--\eqref{eq:reference}, which are formulated for each scenario. Inequalities \eqref{eq:capacity} enforce thermal generating unit capacity limits. Ramping up and down constraints \eqref{eq:ramping} impose limits on the rate of change in generation between consecutive time periods.  Nodal power balance \eqref{eq:balance} ensures that supply equals demand at each bus for every time period and scenario. Transmission line limits \eqref{eq:transmission} enforce thermal capacity limits on transmission lines to prevent overloading. The power flow on line $(i,j)$ is calculated using the DC approximation as the difference in voltage angles divided by the line reactance. Reference bus constraint \eqref{eq:reference} sets the voltage angle at the reference bus to zero for all time periods and scenarios. 

\section{Proposed Surrogate Model}
\label{s3}

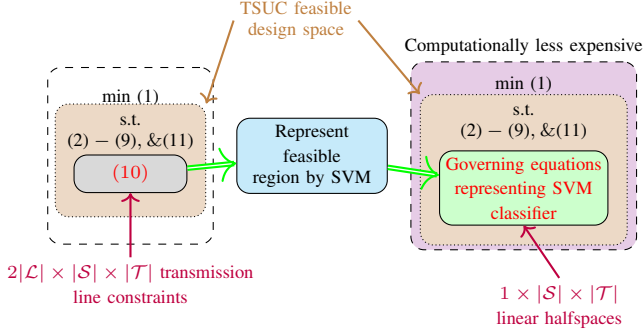
\begin{figure}
    \centering
    \scriptsize
    
    \noindent
    \begin{tikzpicture}   
        \tikzstyle{box} = [draw, rectangle, minimum width=2cm, minimum height=1cm, align=center, rounded corners=5pt]
        \node[box, dashed, minimum width=2.2cm, minimum height=2.35cm, align=center] (box1) {};
        \node[above=-0.6cm of box1.north] {\scriptsize \shortstack{min \(\eqref{eq:objective}\)}};
        \node[draw, densely dotted, fill=brown!30,, align=center, minimum width=2cm, minimum height=1.5cm, above=-2cm of box1.north, rounded corners=5pt] (inner1) {};
        \node [above=-0.7cm of inner1.north] {\scriptsize \shortstack{ s.t. \\ \( \eqref{eq:startup}-\eqref{eq:balance},\&\eqref{eq:reference}\) }};
        \node[draw, rounded corners=5pt, fill=gray!30, minimum width=1.5cm, minimum height=0.5cm, text=red, align=center, below=-1.2cm of box1.south] (inner2) {\scriptsize \shortstack{\( (\ref{eq:transmission}) \) }};
        \node[yshift=-1.7cm, text=purple] (verttext1) {\scriptsize \shortstack {$2|\mathcal{L}|\times|\mathcal{S}|\times  |\mathcal{T}|$ transmission \\ line constraints}};
        \node[box, fill=cyan!20, right=0.3cm of box1] (box2) {\scriptsize \shortstack{Represent \\ feasible \\region by SVM}};
        \node[box, dashed, fill=violet!20, minimum width=3cm, minimum height=2.5cm, align=center, right=0.3cm of box2] (box3) {};
        \node[above=-0.5cm of box3.north] {\scriptsize \shortstack{min \(\eqref{eq:objective}\)}};
         \node[draw, densely dotted, fill=brown!30,, align=center, minimum width=2.75cm, minimum height=2cm, above=-2.45cm of box3.north, rounded corners=5pt] (inner3) {};
        \node [above=-0.7cm of inner3.north] {\scriptsize \shortstack{ s.t. \\ \( \eqref{eq:startup}-\eqref{eq:balance},\&\eqref{eq:reference}\) }};        
        \node[draw, text=red, fill=green!20, text=red, align=center, rounded corners=5pt, above=-1.75cm of inner3.north] (inner4) {\scriptsize \shortstack{Governing equations \\ representing SVM\\ classifier}};
        \node[above=0.0cm of box3.north] {\scriptsize \shortstack {Computationally less expensive}};
        \node[yshift=-2cm, xshift=5.7cm, text=purple] (verttext2) {\scriptsize \shortstack {$1 \times |\mathcal{S}| \times |\mathcal{T}|$ \\linear halfspaces}};
        \node[yshift=1.8cm, xshift=2.2cm, text=brown] (verttext3) {\scriptsize \shortstack {TSUC feasible\\ design space}};
        \draw[->, double, thick, double distance=0.5pt, draw=green, fill=green] (inner2) -- (box2);
        \draw[->, double, thick, double distance=0.5pt, draw=green, fill=green] (box2) -- (inner4);
        \draw[->, thick, draw=purple] (verttext1.north) -- (inner2.south);
        \draw[->, thick, draw=purple] (verttext2.north) -- (inner4.south);
        \draw[->, thick, draw=brown] (verttext3.west) -- (inner1.north east);
        \draw[->, thick, draw=brown] (verttext3.east) -- (inner3.north west);
    \end{tikzpicture}
    \caption{Overview of the proposed approach: training SVM offline and embedding its governing equations into the TSUC problem.}
    \label{Overview of work}
\end{figure}

Fig. \ref{Overview of work} provides an overview of the proposed approach, which replaces a portion of the TSUC feasible region—originally defined by $2|\mathcal{L}|  \times |\mathcal{S}| \times |\mathcal{T}|$ constraints (\ref{eq:transmission})—with a more tractable approximation using $1 \times |\mathcal{S}| \times |\mathcal{T}|$  halfspaces learned from offline simulations. Our approach is grounded in the observation that line flows depend only on net injections and network topology under the DC power flow approximation, not explicitly on the binary commitment variables. This opens the door to modeling line flow feasibility using a supervised learning framework, wherein the feasibility of a given operating point can be evaluated from its feature representation. In this section, we provide a detailed justification for this surrogate modeling approach and develop a theoretical foundation for using linear SVMs to classify feasible versus infeasible operating conditions.

\subsection{Justification for Data-driven Modeling of Network Constraints Using DCOPF}

In the TSUC formulation, the network constraints \eqref{eq:transmission} are derived from the DC power flow approximation. Although these constraints appear in TSUC, we argue that they can be modeled independently using a surrogate learned from data generated via DCOPF, thus decoupling them from the binary decision variables in the TSUC problem. Below, we justify this approach.

\subsubsection{Decoupling Network Constraints from Unit Commitment Decisions}

Under the DC power flow approximation, the power flow on line $(i,j)$ 
denoted as $f_{ij,s,t}$ is calculated as:
\begin{equation}
f_{ij,s,t} = \frac{\theta_{i,s,t} - \theta_{j,s,t}}{x_{ij}}
\label{eq:line_flow_dc}
\end{equation}
These voltage angles are determined from the system of linear equations formed by the power balance constraints at each bus:

\begin{equation}
\begin{split}
\sum_{g \in \mathcal{G}_i} p_{g,s,t} + P^{wind}_{s,t}
- \sum_{j \in \mathcal{N}_i} \frac{\theta_{i,s,t} - \theta_{j,s,t}}{x_{ij}} \\
= D_{i,s,t}, \quad \forall i \in \mathcal{B},\; t \in \mathcal{T},\; s \in |\mathcal{S}|
\end{split}
\label{eq:power_balance}
\end{equation}

The key insight is that the voltage angles $\theta_{i,s,t}$ depend only on the net power injections (generation minus demand) and the network topology. They do not depend explicitly on the UC binary variables $u_{g,t} \in \{0,1\}$, which govern the ON/OFF status of generators.

While the UC variables constrain the feasible space of generator outputs via constraints such as \eqref{eq:capacity}, the line flows are only functions of the actual dispatched power. Thus, for any given realization of net injections $\{p_{g,s,t}, P^{wind}_{s,t}, D_{i,s,t}\}$, the resulting line flows can be computed independently of how those injections were determined (i.e., with or without the presence of binary UC logic).

\subsubsection{Surrogate Modeling via DCOPF Data}

Let $\boldsymbol{\phi}_{t,s} = \left[\boldsymbol{\mu}_s, \boldsymbol{\sigma}_s, \mathbf{p}_{s,t} \right]^\top$ denote the feature vector representing uncertainty parameters and the base dispatch at time $t$ and scenario $s$, where $\boldsymbol{\mu}_s = [\mu_{1,s}, \mu_{2,s}, \ldots, \mu_{n,s}]^\top$ is the mean vector of uncertain parameters, $\boldsymbol{\sigma}_s = [\sigma_{1,s}, \sigma_{2,s}, \ldots, \sigma_{n,s}]^\top$ is the standard deviation of uncertain parameters, and $\mathbf{p}_{s,t} = [p_{1,s,t}, p_{2,s,t}, \ldots, p_{|\mathcal{G}|,s,t}]^\top$ is the dispatch vector. 
We aim to learn a surrogate model $\hat{f}_{ij}(\boldsymbol{\phi}) \approx f_{ij,s,t}$ from data. To train such a model, we generate synthetic data using DCOPF simulations under various renewable generation and load scenarios.
The DCOPF equations are given below.

\begin{align}
\min_{p_g, \theta} \quad & \sum_{g \in \mathcal{G}} C_g(p_g) \\
\text{s.t.} \quad
& \sum_{g \in \mathcal{G}_i} p_{g} + P^{wind} - \sum_{j \in \mathcal{N}_i} \frac{\theta_i - \theta_j}{x_{ij}} = D_i, \quad \forall i \in \mathcal{B}  \label{eq:dcopf_balance} \\
& P_{g}^{\min} \leq p_{g} \leq P_{g}^{\max}, \quad \forall g \in \mathcal{G} \label{eq:dcopf_gen_limits} \\
& \left| \frac{\theta_i - \theta_j}{x_{ij}} \right| \leq \bar{f}_{ij}, \quad \forall (i,j) \in \mathcal{L} \label{eq:dcopf_line_limits} \\
& \theta_{\text{ref}} = 0 \label{eq:dcopf_theta_ref}
\end{align}

The DCOPF assumes that all generators are online and dispatched optimally, subject to physical constraints. As a result, the data generated from DCOPF captures the relationship between power injections and line flows under the DC approximation.

\subsubsection{Theoretical Justification}

The mathematical foundation for this decoupling approach rests on the linearity of the DC power flow equations. Given the network admittance matrix $\mathbf{Y}$ and the vector of net injections $\mathbf{p}$, the voltage angles are determined by:
\begin{equation}
\boldsymbol{\theta} = \mathbf{Y}^{-1} \mathbf{p}
\label{eq:voltage_angles}
\end{equation}

Since this relationship is linear, the line flows are also linear functions of the net injections. The UC variables only affect generators that can contribute to the net injections, not the physical relationship between injections and flows.
Therefore, under the linear DC power flow model, the network constraints depend only on power injections and network parameters and are independent of UC decisions. This validates using a data-driven surrogate model trained on DCOPF data to approximate line flow constraints in the TSUC problem. 

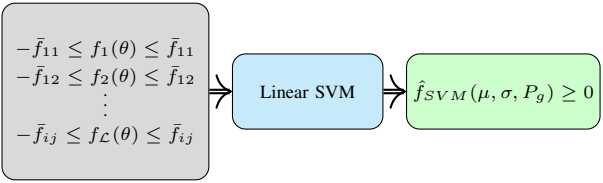
\begin{figure}
    \centering
    \scriptsize
    
    \noindent
    \begin{tikzpicture}   
        \tikzstyle{box} = [draw, rectangle, minimum width=2cm, minimum height=1cm, align=center, rounded corners=5pt]
        \node[box, fill=gray!30, minimum width=2.75cm, minimum height=2.35cm, align=center] (box1) {};
        \node {\scriptsize \shortstack{$-\bar{f}_{11} \leq f_1(\theta) \leq \bar{f}_{11}$ \\ $-\bar{f}_{12} \leq f_2(\theta) \leq \bar{f}_{12}$ \\ . \\.\\.\\ $-\bar{f}_{ij} \leq f_\mathcal{L}(\theta) \leq \bar{f}_{ij}$}};
        \node[box, fill=cyan!20, right=0.3cm of box1] (box2) {\scriptsize \shortstack{Linear SVM}};
        \node[box, fill=green!20, right=0.3cm of box2] (box3) {\scriptsize \shortstack{$ \hat{f}_{SVM}(\mu,\sigma,P_g) \geq 0$}};
        \draw[->, double, thick, double distance=0.5pt] (box1) -- (box2);
        \draw[->, double, thick, double distance=0.5pt] (box2) -- (box3);
    \end{tikzpicture}
    \caption{Single surrogate halfspace $f_{SVM}$ constructed from governing equations of SVM to replace all $2|\mathcal{L}|$ transmission line constraints.}
    \label{SVM surrogate}
\end{figure}

\textbf{Implication:} As illustrated in Fig.~\ref{SVM surrogate}, all $2|\mathcal{L}|$ transmission line flow constraints \eqref{eq:transmission} can be replaced by a single surrogate halfspace $\hat{f}_{\text{SVM}}(\mu, \sigma, P_g) \geq 0$, constructed using the governing equations of an SVM trained on features derived from the DCOPF. The surrogate constraint allows for computationally less expensive and decoupled modeling of transmission feasibility within a data-driven TSUC framework. This approach reduces computational complexity while preserving the essential physics of power flow constraints.

\subsection{Justification and Feature Vector Construction for Linear Surrogate Modeling}

The use of linear SVMs as surrogate models for transmission line flow constraints is justified by both mathematical properties of the DC power flow model and practical observations from system operations. Under the DC approximation, the transmission feasibility region is globally a polyhedron with $2|\mathcal{L}|$ faces, where each inequality corresponds to an upper or lower thermal limit on a line. Enforcing these constraints directly in a TSUC formulation leads to $2|\mathcal{L}| \times |\mathcal{S}| \times |\mathcal{T}|$ inequalities across all scenarios and time periods, significantly hindering scalability.

The feasibility of linear SVM approximation relies on the \textit{effective dimensionality reduction} and \textit{approximate linear separability} principles within typical power system operational envelopes. Specifically, the following conditions ensure that a single affine hyperplane can approximate the feasibility boundary:

\begin{enumerate}
    \item \textbf{Restricted Operational Envelope:} Power systems operate within a compact region around nominal loading (typically 70--130\% of base case). Within this restricted set $\Omega$, feasibility is governed by a limited number of active facets, so the feasible/infeasible boundary potentially coincides with a single supporting halfspace, eliminating the need to represent the entire polyhedron \cite{baker2019joint,10665966}.
    
    \item \textbf{Sparse Active Constraints:} Although the polyhedral feasibility region has $2|\mathcal{L}|$ faces, empirical evidence shows that only a small fraction of line flow limits bind simultaneously. This sparse binding pattern means that the feasible boundary is often governed by a small subset of constraints, enabling effective local approximation by a single supporting halfspace.
    
    \item \textbf{DC Approximation Linearization:} The DC power flow model already linearizes the relationship between generation dispatch and line flows, so the feasible set is polyhedral. Although globally this polyhedron may have many facets, within the restricted envelope $\Omega$ only a small subset is relevant. In such settings, the feasible/infeasible boundary can be represented by a linear SVM hyperplane.

    \item \textbf{Feature-Space Separability:} Feature-space separability is central to justifying a linear SVM surrogate for replacing polyhedral transmission constraints. When feasible and infeasible operating points are embedded in a feature space defined by stochastic parameters and controllable generation decisions, $\boldsymbol{\phi} = [\mu, \sigma, p_g]^\top$, the two classes cluster on opposite sides of an approximate hyperplane. Formally, there exists $(\mathbf{w}, b)$ such that
    \begin{equation}
    y_i (\mathbf{w}^T \phi_i + b) \geq \gamma - \epsilon, \quad \forall i,
    \end{equation}
    where $\gamma > 0$ denotes a margin, and $\epsilon$ is a small tolerance. 
    
    The margin is defined as the minimum distance between the separating hyperplane and the closest data points from either class (the support vectors). A larger margin indicates a more robust separation, as it improves generalization and reduces sensitivity to perturbations. 
    
    The determination of $(\mathbf{w}, b)$ arises from an optimization problem whose objective is to maximize this margin subject to the above inequality constraints. This optimization effectively balances two goals: (i) maximizing the distance between feasible and infeasible operating points, and (ii) permitting a small tolerance $\epsilon$ that accounts for approximate rather than strict separability. In this way, even if global linear separability is not attainable, the learned hyperplane captures the dominant feasibility boundary in the high-density operational region relevant to power system operations.

    In this study, the maximum separating margin, evaluated with a small tolerance ($10^{-9}$), is obtained as 1.20 for the IEEE 57-bus system and 1.25 for the IEEE 118-bus system. Thus, the separability property explains why a \emph{single learned hyperplane} can reliably approximate the boundary of feasible operation, reducing  $2|\mathcal{L}|$ network constraints to one surrogate inequality per scenario and time step while maintaining security guarantees.
    
    It is also important to emphasize that the observed separability is closely tied to the choice of feature vectors. In other words, linear separability may not solely an inherent property of the DCOPF formulation, but is facilitated by embedding operating points into an appropriately defined feature space. This observation is conceptually related to the well-known \emph{kernel trick} in support vector machines: while data in the $(x_1, x_2)$ plane may not be linearly separable, augmenting the representation to $(x_1, x_2, x_1^2+x_2^2)$ can achieve separability. In a similar manner, the constructed feature vector in this study provides a linear hyperplane to approximate the feasibility boundary with high accuracy.

    \item \textbf{Statistical Robustness:} Although strict global separability may not hold, empirical misclassification rates within the sampled operating envelope are negligible. More importantly, misclassifications are not equally critical: classifying an infeasible point as feasible (false positive) risks violating transmission limits, whereas classifying a feasible point as infeasible (false negative) is merely conservative. To reflect this asymmetry, we employ weighted classification, assigning a higher penalty to misclassifying infeasible points. This adjustment shifts the separating hyperplane to favor security, substantially reducing the likelihood of false positives while tolerating a small number of false negatives. In practice, the surrogate hyperplane thus errs on the side of caution, analogous to security-constrained and chance-constrained OPF, ensuring that reliability is maintained even in the presence of minor statistical errors.
\end{enumerate}

\textbf{Mathematical Formulation:} Let $\mathcal{F} = \{\mathbf{x} : \mathbf{A}\mathbf{x} \leq \mathbf{b}\}$ represent the polyhedral feasible region with $2|\mathcal{L}|$ constraints. For a given operational envelope $\Omega$, the restricted feasible region is $\mathcal{F}_\Omega = \mathcal{F} \cap \Omega$. The effective boundary of $\mathcal{F}_\Omega$ can be approximated by:
\begin{equation}
\mathbf{w}^T \mathbf{x} + b_{\text{eff}} \geq 0
\end{equation}
where $\mathbf{w}$ and $b_{\text{eff}}$ are learned from operational data, capturing the statistically dominant feasibility boundary within the operational envelope.

A linear SVM trained on labeled operating points (feasible/infeasible) generated from DCOPF simulations leverages this property. Rather than reconstructing the entire polyhedron, the SVM identifies the dominant separating hyperplane in the chosen feature space:
\begin{equation}
    \hat{f}_{\text{SVM}}(\boldsymbol{\phi}_{s,t}) \geq 0,
    \label{eq:fsvm}
\end{equation}
where $\boldsymbol{\phi}_{s,t}$ denotes the feature vector at scenario $s$ and time $t$. This surrogate reduces the dimensionality of the transmission constraints from $2|\mathcal{L}| \times |\mathcal{S}| \times |\mathcal{T}|$ inequalities to a compact form of $1 \times |\mathcal{S}| \times |\mathcal{T}|$. Indeed, the optimization feasible design space will be changed from \[\label{eq:original}
\mathcal{F}_{\text{original}} = \{ \text{Constraints } (\ref{eq:startup}) \text{--} (\ref{eq:reference}) \}\] to  \[\mathcal{F}_{\text{proposed}} = \{ \text{Constraints } (\ref{eq:startup}) \text{--} (\ref{eq:balance}),  (\ref{eq:reference}) \cup (\ref{eq:fsvm})\}.\]

In this formulation, the SVM surrogate acts as a data-driven approximation of the original polyhedral network constraints, thereby retaining feasibility information while significantly reducing the number of explicit constraints in the problem.

The effectiveness of this approximation depends critically on the definition of the feature vector $\boldsymbol{\phi}_{s,t}$. Guided by theoretical analysis, the feature vector must capture both the stochastic properties of uncertainty and the controllable generation schedule. A practical construction is:
\[
\boldsymbol{\phi}_{s,t} = \left[\mu_{1,s,t}, \ldots, \mu_{n,s,t}, \sigma_{1,s,t}, \ldots, \sigma_{n,s,t}, \mathbf{p}_{g,s,t} \right]^\top,
\]
where $\mu_{i,s,t}$ and $\sigma_{i,s,t}$ denote the mean and standard deviation of uncertain parameters (e.g., wind generation and load variations), and $\mathbf{p}_{g,s,t}$ represents the baseline generator dispatch. By embedding these features, the surrogate model faithfully reflects the influence of uncertainty and controllable decisions on line feasibility.

Moreover, because SVMs maximize the separating margin, the resulting surrogate inequality is naturally conservative, thereby reducing the risk of infeasible operating points being misclassified as feasible. By integrating this surrogate representation with carefully constructed features, transmission feasibility constraints are enforced through a single governing inequality, substantially reducing computational complexity while preserving fidelity to the underlying physics.

\begin{remark}[Interpretation]
The linear SVM surrogate does not replace the global polyhedron by a single halfspace. Instead, it learns the \emph{supporting cut} that explains feasibility within the operating envelope actually encountered. This is justified because only a few network constraints are active in practice, and their intersection is approximately affine in the chosen feature space, enabling effective separation through a hyperplane. The surrogate hyperplane, therefore, provides a data-driven yet physically consistent reduction of network constraints.
\end{remark}

\subsection{Integration into TSUC Formulation}
To operationalize the proposed surrogate modeling approach within the TSUC framework, the decision boundary learned through the SVM classifier must be seamlessly embedded as a replacement for the explicit transmission network constraints. Leveraging the theoretical insights into the convex polyhedral structure of the feasible region, the SVM-generated hyperplane (the inequality constraint) defined by the linear SVM---parameterized by the learned weight vector and bias term---can be directly incorporated into the TSUC formulation as a surrogate feasibility constraint. This substitution effectively reduces the dimensionality of the network constraint set from $2|\mathcal{L}| \times |\mathcal{S}| \times |\mathcal{T}|$ to $1 \times |\mathcal{S}| \times |\mathcal{T}|$, thereby enhancing scalability without compromising constraint satisfaction. 

The SVM-based surrogate constraint, derived from the governing equations of an offline-trained SVM \cite{burges1998tutorial}, is expressed as:
\begin{align}
\mathbf{w}^T\boldsymbol{\bm{\phi}}_{t,s} + b \geq 0, \quad \forall s \in \mathcal{S}, \forall t \in \mathcal{T} \label{eq:svm_constraint}
\end{align}
where $\mathbf{w}$ and $b$ are the weight vector and bias learned from DCOPF simulation data. These learned halfspaces directly replace the original line flow constraints in the TSUC formulation, yielding the modified TSUC problem:
\begin{align}
& \min \eqref{eq:objective} \\
& \text{s.t.} \nonumber \\
&    \eqref{eq:startup}-\eqref{eq:balance}, \& \eqref{eq:reference}, \nonumber \\
&    \text{SVM generated surrogate feasibility constraint} \eqref{eq:svm_constraint} \nonumber
\end{align}

This substitution is not only computationally efficient and scalable to large systems and multiple scenarios, but also grounded in a solid theoretical foundation: the polyhedral structure of the SVM margin offers a mathematically justified approximation of the feasible region. Furthermore, the inherent margin maximization of SVM ensures a conservative approximation, reducing the risk of feasibility violations. This modification enables the TSUC formulation to bypass infeasible regions in the solution space while retaining key operational constraints, offering a practical, theoretically sound, scalable solution framework.

\begin{figure}[htbp]
\centering
\scriptsize

\resizebox{0.45\textwidth}{!}{
\begin{tikzpicture}[
    scale=0.5, 
    node distance=0.65cm and 0.5cm,
    input/.style={rectangle, rounded corners, minimum width=1.25cm, minimum height=1cm, text centered, draw=black, fill=blue!20, align=center},
    process/.style={rectangle, minimum width=2cm, minimum height=1cm, text centered, draw=black, fill=orange!20, align=center},
    decision/.style={diamond, minimum width=2cm, minimum height=1cm, text centered, draw=black, fill=green!20, align=center},
    arrow/.style={line width=0.6mm, ->, >=Stealth, align=center} 
]
\node (data) [input] {DCOPF Simulations\\ (Wind + Load Scenarios)};
\node (train) [process, below=of data] {SVM Training\\ (Feasible vs Infeasible)};
\node (svm) [decision, below=of train] {SVM Hyperplane\\ Line Flow Constraints \\ Surrogate Model };
\node (tsuc) [process, right=4cm of svm] {TSUC Optimization\\ with Surrogate Model \\ (SVM Hyperplane)};
\node (solution) [input, below=of tsuc] {Computationally Less \\ Expensive TSUC};

\draw [arrow] (data) -- (train);
\draw [arrow] (train) -- (svm);
\draw [arrow] (svm) -- node[above] {Replace \\ Line Flow Constraints} (tsuc);
\draw [arrow] (tsuc) -- (solution);

\node at ([xshift=0.1cm, yshift=0.5cm]data.north west) [anchor=south west] {\textbf{\large Offline Phase}};
\node at ([xshift=0.2cm, yshift=0.5cm]tsuc.north east) [anchor=south east] {\textbf{\large Online Phase}};
\end{tikzpicture}
}
\caption{Workflow of proposed SVM surrogate TSUC modeling framework.}
\label{fig:surrogate_workflow}
\end{figure}
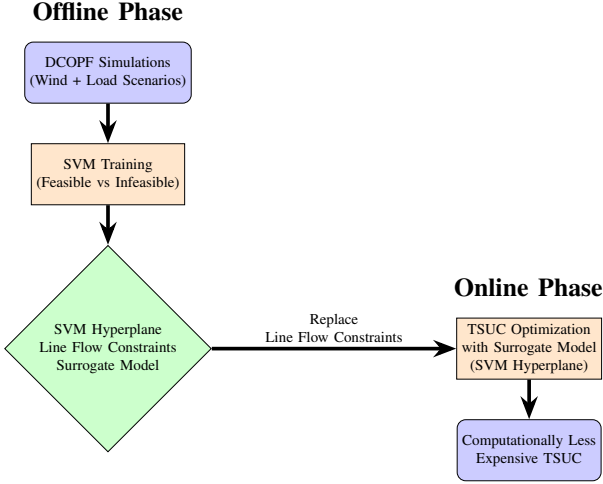

The workflow of the proposed SVM-based surrogate modeling approach for TSUC is depicted in Fig.~\ref{fig:surrogate_workflow}. It consists of an offline phase where DCOPF simulations generate training data for SVM classification of feasible and infeasible operating points, followed by an online phase where the resulting SVM halfspace is a surrogate constraint to replace transmission line constraints in the TSUC optimization problem.

\section{Simulation Results}
\label{2}
The simulation results were generated using the IEEE 57-bus and 118-bus test systems. All data related to the TSUC problem—including wind uncertainty scenarios, load variations, and DCOPF feasibility labels—were produced in \textsc{Matlab} using the \textsc{CPLEX} solver along with the \textsc{MATPOWER} and \textsc{YALMIP} toolboxes. The subsequent training, testing, and hyperplane extraction via SVM were performed in \textsc{Python}. All computations were carried out on a personal computer equipped with an Intel(R) Xeon(R) CPU E5-1630 v3 @ 3.70\,GHz, 16\,GB of RAM, and a 64-bit operating system.


\subsection{Data Generation}
The data generation process uses a Monte Carlo simulation approach to create a comprehensive dataset for training machine learning models to provide a surrogate model for line flow constraints in the DCOPF problem. This section outlines the method for generating synthetic datasets for solving the DCOPF problem under wind and load uncertainty in the 57- and 118-bus systems. The simulation framework systematically introduces variability in wind generation and load demand, solves the DCOPF for each scenario, and records the resulting optimal generation dispatch, line power flows, and feasibility status. 

We used a two-step data generation approach to ensure that the generated dataset adequately captures a representative range of operating conditions. First, simulations were performed with line flow constraints, and all resulting operating points were labeled as feasible. Second, line flow constraints were relaxed, and each operating point was individually evaluated for feasibility based on the original constraints, allowing for the identification and labeling of infeasible cases. This iterative process was repeated until a balanced dataset—comprising comparable numbers of feasible and infeasible samples—was achieved. 

The data generation procedure was carried out through the following detailed explanation: Wind generation units are integrated at buses 12, 25, 33, 45, and 53 in the 57-bus system, and at buses 9, 31, 54, 90, and 100 in the 118-bus system. The mean and standard deviation of wind generation at each wind bus are denoted, and for each simulation sample, these values are independently drawn from predefined intervals as follows:
\begin{equation}
\mu_s \sim \mathcal{U}(\mu_{\text{lower},s}, \mu_{\text{upper},s}), \quad
\sigma_s \sim \mathcal{U}(\sigma_{\text{lower},s}, \sigma_{\text{upper},s})
\end{equation}

To explore the entire domain of wind uncertainty, a parameter $z_s$ is introduced, varying from -4 to 4 in increments of 0.1. This approach captures a wide range of probabilistic scenarios corresponding to standard deviations around the mean. The wind power realization at each wind bus is then computed as:
\begin{equation}
P^{wind}_s(z) = \mu_s + z_s\cdot \sigma_s
\end{equation}

To account for stochastic variations in load, each bus demand is independently perturbed within the range of 70\% to 130\% of its nominal value, using samples drawn from the uniform distribution.

After executing the simulation for each scenario, the resulting line flows are compared against their respective thermal limits to determine feasibility. If all line flows remain within their allowable limits, the scenario is labeled as feasible; otherwise, it is classified as infeasible. The respective means and standard deviations of wind generation, along with the active power outputs of conventional generators and their corresponding feasibility labels, are used as input features for training the SVM. The generated dataset is partitioned into 80\% for training and 20\% for testing to evaluate the performance of the trained SVM. Feature scaling is applied to normalize the input variables before training. The classifier is trained with a linear kernel, and hyperparameters are tuned using grid search and cross-validation.

\subsection{Results}
The SVM model with a linear kernel exhibited strong classification performance in distinguishing between feasible and infeasible scenarios for both the 57-bus and 118-bus test systems. The testing accuracies achieved were 99.72\% for the 57-bus system and 99.88\% for the 118-bus system. The corresponding confusion matrices for each case study are presented in Fig.~\ref{fig:svm_confusion_matrices}.

\begin{figure}[htbp]
    \centering
    \scriptsize

    \begin{subfigure}[b]{0.241\textwidth}
        \centering
        \begin{tikzpicture}[scale=0.7]
            \draw[thick] (0,0) rectangle (4,4);
            \draw[thick] (2,0) -- (2,4);
            \draw[thick] (0,2) -- (4,2);

            \node[font=\scriptsize] at (1,4.3) {\textbf{Infeasible}};
            \node[font=\scriptsize] at (3,4.3) {\textbf{Feasible}};
            \node[font=\scriptsize, rotate=90] at (-0.25,3) {\textbf{Infeasible}};
            \node[font=\scriptsize, rotate=90] at (-0.25,1) {\textbf{Feasible}};

            \node[fill=green!20, rounded corners, minimum width=1cm, minimum height=1cm] at (1,3) {\scriptsize 710};
            \node[fill=orange!30, rounded corners, minimum width=1cm, minimum height=1cm] at (3,3) {\scriptsize 3};
            \node[fill=red!30, rounded corners, minimum width=1cm, minimum height=1cm] at (1,1) {\scriptsize 1};
            \node[fill=green!20, rounded corners, minimum width=1cm, minimum height=1cm] at (3,1) {\scriptsize 706};

            \node[font=\scriptsize] at (2,-0.2) {\textbf{Predicted Class}};
            \node[font=\scriptsize, rotate=90] at (-.75,2) {\textbf{Actual Class}};
        \end{tikzpicture}
        \caption{\scriptsize IEEE 57-bus system}
    \end{subfigure}
    \hfill
    \begin{subfigure}[b]{0.241\textwidth}
        \centering
        \begin{tikzpicture} [scale=0.7]
            \draw[thick] (0,0) rectangle (4,4);
            \draw[thick] (2,0) -- (2,4);
            \draw[thick] (0,2) -- (4,2);

            \node[font=\scriptsize] at (1,4.3) {\textbf{Infeasible}};
            \node[font=\scriptsize] at (3,4.3) {\textbf{Feasible}};
            \node[font=\scriptsize, rotate=90] at (-0.25,3) {\textbf{Infeasible}};
            \node[font=\scriptsize, rotate=90] at (-0.25,1) {\textbf{Feasible}};

            \node[fill=green!20, rounded corners, minimum width=1cm, minimum height=1cm] at (1,3) {\scriptsize 788};
            \node[fill=orange!30, rounded corners, minimum width=1cm, minimum height=1cm] at (3,3) {\scriptsize 2};
            \node[fill=red!30, rounded corners, minimum width=1cm, minimum height=1cm] at (1,1) {\scriptsize 0};
            \node[fill=green!20, rounded corners, minimum width=1cm, minimum height=1cm] at (3,1) {\scriptsize 810};

            \node[font=\scriptsize] at (2,-0.2) {\textbf{Predicted Class}};
            \node[font=\scriptsize, rotate=90] at (-.75,2) {\textbf{Actual Class}};
        \end{tikzpicture}
        \caption{\scriptsize IEEE 118-bus system}
    \end{subfigure}

    \caption{Confusion matrices of linear SVM classifier for IEEE 57- and 118-bus systems. Green: correct classifications; Red/Orange: misclassifications.}
    \label{fig:svm_confusion_matrices}
\end{figure}

The linear SVM classifier demonstrates excellent performance for both test systems. For the 57-bus system, the classifier correctly identified 710 out of 713 infeasible scenarios with only three false positives. It achieved near-perfect classification of feasible scenarios with 706 out of 707 correctly identified and only one false negative. The 118-bus system shows even better performance, with the classifier correctly identifying 788 out of 790 infeasible scenarios with only two false positives, and perfectly classifying all 810 feasible scenarios with no false negative. This performance suggests well-learned decision boundaries in both systems, though the slight asymmetry in misclassification—primarily on the infeasible side—may reflect underlying complexities in the feasible boundary region of the DCOPF solution space. In practical applications, especially those involving power system operations, minimizing false positives is essential to maintaining system security and avoiding constraint violations. Given its extremely low false positive rate (only three cases for the 57-bus system and two for the 118-bus system), the classifier is highly suitable for integrating into the TSUC optimization framework as a surrogate model for network constraints.

The hyperplanes must be extracted from the linear SVM classifiers trained on the 57-bus and 118-bus systems and subsequently integrated into the TSUC optimization framework. This extraction process involves obtaining the decision boundary parameters (weight vectors and bias terms) that define the separating hyperplane in the feature space. The resulting surrogate hyperplanes for the 57- and 118-bus systems, obtained through this extraction process, are presented in \eqref{57bus equation} and \eqref{118bus equation}, respectively. The coefficients corresponding to the other power generation units were nearly zero and, therefore, negligible.

\begin{align}
&\hat{f}_{\text{SVM-57-bus}}(\mu_1, \ldots, \mu_5, \sigma_1, \ldots, \sigma_5, Pg_1, Pg_3, Pg_5, Pg_7) = \nonumber \\ \nonumber
& \quad + 2.07 \cdot \mu_1 + 4.87 \cdot \mu_2 + 1.24 \cdot \mu_3 + 1.46 \cdot \mu_4 - 3.12 \cdot \mu_5 \\ \nonumber
&\quad - 7.84 \cdot \sigma_1 - 0.27 \cdot \sigma_2 - 0.94 \cdot \sigma_3 - 1.78 \cdot \sigma_4 + 0.08 \cdot \sigma_5 & \\ \nonumber
&\quad - 0.0092 \cdot Pg_1 - 0.016 \cdot Pg_3 - 16.97 \cdot Pg_5 - 0.008 \cdot Pg_7 \\
&\quad + 24.34
\label{57bus equation}
\end{align}
\begin{align}
&\hat{f}_{\text{SVM-118-bus}}(\mu_1, \ldots, \mu_5, \sigma_1, \ldots, \sigma_5, Pg_1, \dots , Pg_{54}) = \nonumber \\ \nonumber
& \quad + 2.69 \cdot \mu_1 - 2.24 \cdot \mu_2 - 2.31 \cdot \mu_3 + 0.4 \cdot \mu_4 + 1.54 \cdot \mu_5 \\ \nonumber
&\quad - 3.32 \cdot \sigma_1 + 1.71 \cdot \sigma_2 - 1.67 \cdot \sigma_3 - 2.22 \cdot \sigma_4 - 0.36 \cdot \sigma_5 & \\ \nonumber
&\quad - 0.02 \cdot Pg_6 + 2.23 \cdot Pg_{14} + 1.09 \cdot Pg_{22} + 0.05 \cdot Pg_{28} \\ \nonumber
&\quad - 0.96 \cdot Pg_{30} - 0.21 \cdot Pg_{39} - 1.63 \cdot Pg_{40} - 1.39 \cdot Pg_{45} \\
& \quad + 35.4 \cdot Pg_{46} + 0.19\cdot Pg_{51} + 20.4
\label{118bus equation}
\end{align}

\begin{figure}[htbp]
    \centering
    \scriptsize
    
    \begin{subfigure}[b]{0.241\textwidth}
        \centering
        \includegraphics[width=\linewidth]{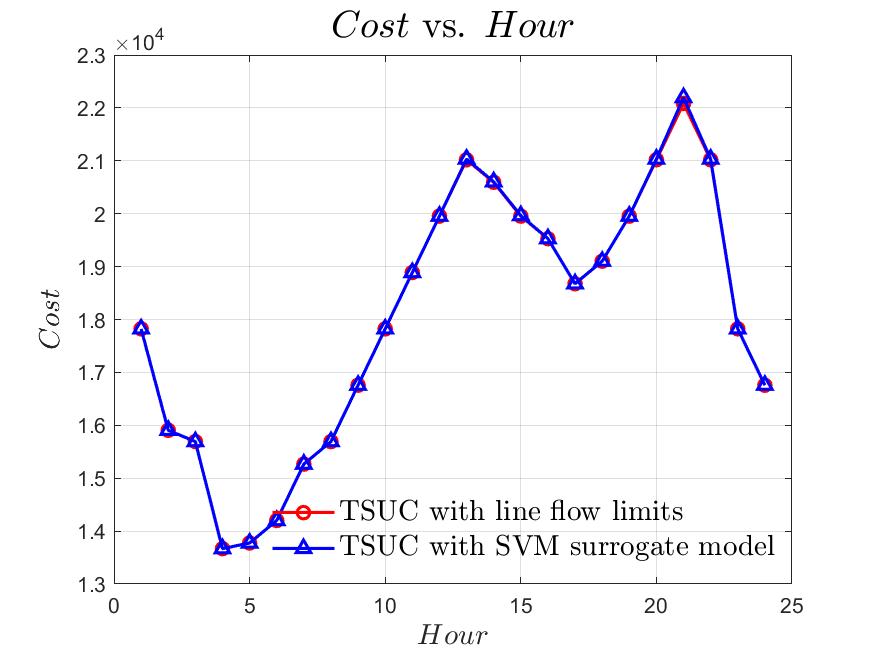}
        \caption{}
        \label{fig:sub1}
    \end{subfigure}
    \hfill
    \begin{subfigure}[b]{0.241\textwidth}
        \centering
        \includegraphics[width=\linewidth]{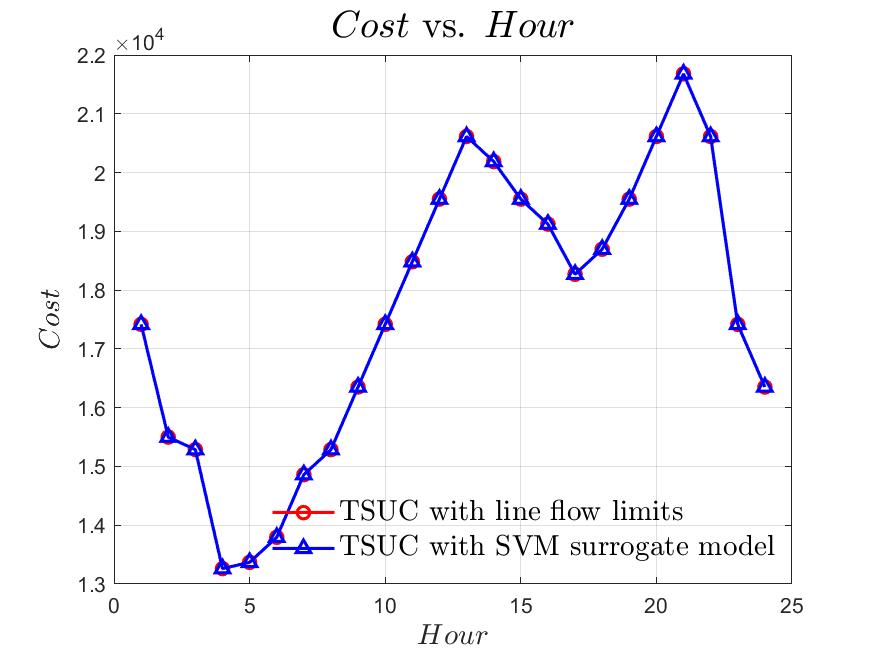}
        \caption{}
        \label{fig:sub2}
    \end{subfigure}
    \caption{Generation costs for two samples of 57-bus system: (a) sample 1, (b) sample 2.}
    \label{a}
\end{figure}

\begin{figure}[htbp]
    \centering
    \begin{subfigure}[b]{0.241\textwidth}
        \centering
        \includegraphics[width=\linewidth]{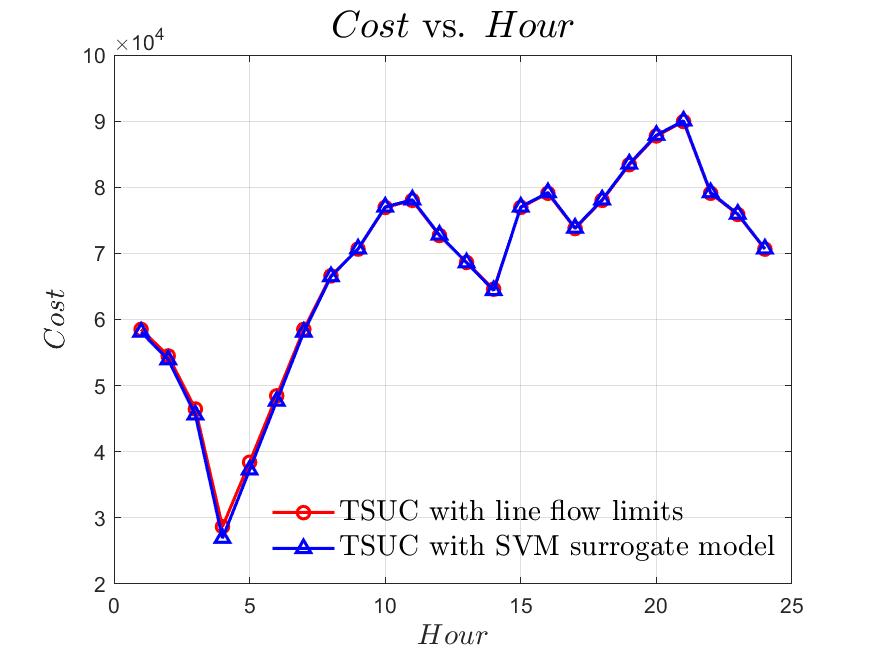}
        \caption{}
        \label{fig:sub11}
    \end{subfigure}
    \hfill
    \begin{subfigure}[b]{0.241\textwidth}
        \centering
        \includegraphics[width=\linewidth]{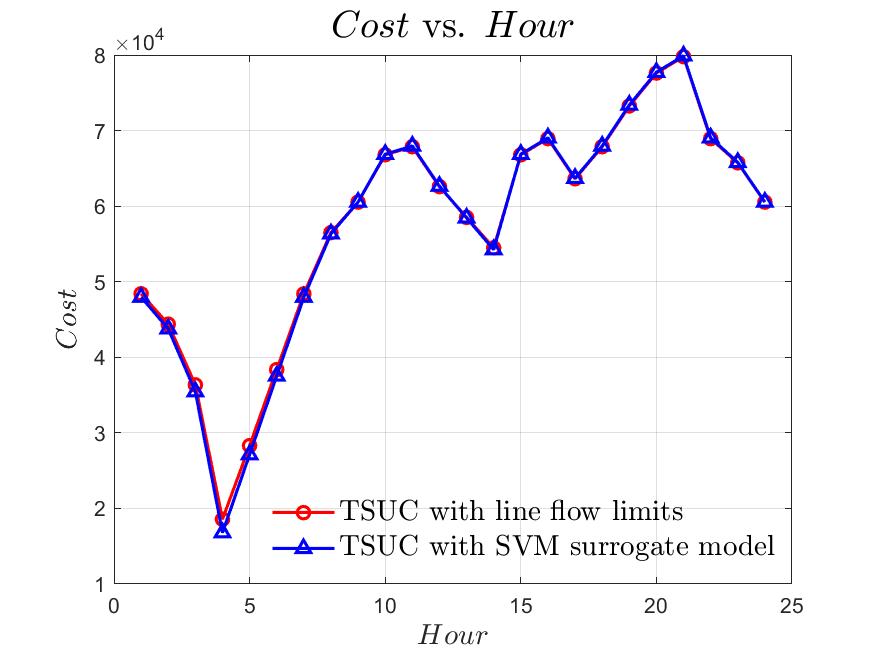}
        \caption{}
        \label{fig:sub21}
    \end{subfigure}
    \caption{Generation costs for two samples of 118-bus system: (a) sample 1, (b) sample 2.}
    \label{b}
\end{figure}

Two randomly selected samples from the testing dataset are presented for each case study to illustrate the TSUC objective function. The thermal unit generation costs for the IEEE 57-bus and 118-bus systems are shown in Fig.~\ref{a} and ~\ref{b}, respectively. For the 57-bus system, the percentage error in the unit commitment expected cost between the SVM-surrogate TSUC and the traditional TSUC is 0.04\% for sample 1 and nearly zero for sample 2, demonstrating accuracy consistent with the confusion matrix analysis. For the 118-bus system, the corresponding errors are 0.3\% and 0.4\%, respectively.

Table~\ref{tab:line_surrogate} highlights the substantial reduction in the number of line flow constraints achieved by replacing all explicit transmission line flow limits with a single SVM-based surrogate constraint in the TSUC formulation. In the original problem, each scenario and time period is associated with hundreds of individual line flow constraints—160 for the 57-bus system and 372 for the 118-bus system. The surrogate-based approach consolidates these into a single linear constraint per sample and hour, representing a significant reduction in dimensionality. 

Table~\ref{tab:comp_time_surrogate} summarizes the computational time statistics for solving the TSUC problem under two configurations: (i) with explicit line flow constraints and (ii) with the SVM-based surrogate model. Simulations were conducted using all samples within the SVM testing dataset. The minimum, maximum, and average time saving percentages are reported in the table. The surrogate model led to an average reduction in computational time of approximately 46\% for the 57-bus system and 31\% for the 118-bus system. 

\begin{table}[t]
    \centering   
    \caption{Reduction in number of network constraints using proposed SVM-based surrogate model in TSUC formulation ($\text{\# line flow constraints} \times \text{\# samples} \times \text{\# hours}$).}
    \begin{tabularx}{\columnwidth}{l *{3}{>{\centering\arraybackslash}X}}
        \toprule
        \textbf{System} & \textbf{Traditional TSUC} & \textbf{Proposed model} & \textbf{Dimensionality Reduction (\%)} \\
        \midrule
        57-bus  & $160\times20\times24=76,800$ & $1\times20\times24=480$ & 99.38 \\
        118-bus & $372\times50\times24=446,400$ & $1\times50\times24=1,200$ & 99.73 \\
        \bottomrule
    \end{tabularx}
    \label{tab:line_surrogate}
\end{table}

\begin{table}[t]
    \centering
    \scriptsize 
    
    \caption{TSUC computational time reduction by proposed SVM-based surrogate model}
    \label{tab:comp_time_surrogate}
    \begin{tabularx}{\columnwidth}{l *{3}{>{\centering\arraybackslash}X}}
        \toprule
        \textbf{Test Systems} & \textbf{Min (\%)} & \textbf{Max (\%)} & \textbf{Average (\%)} \\
        \midrule
        IEEE 57-bus  & 43 & 50 & 46 \\
        IEEE 118-bus & 26 & 39 & 31 \\
        \bottomrule
    \end{tabularx}
\end{table}

\begin{figure}[htbp]
    \centering
    \begin{subfigure}[b]{0.241\textwidth}
        \centering
        \includegraphics[width=\linewidth]{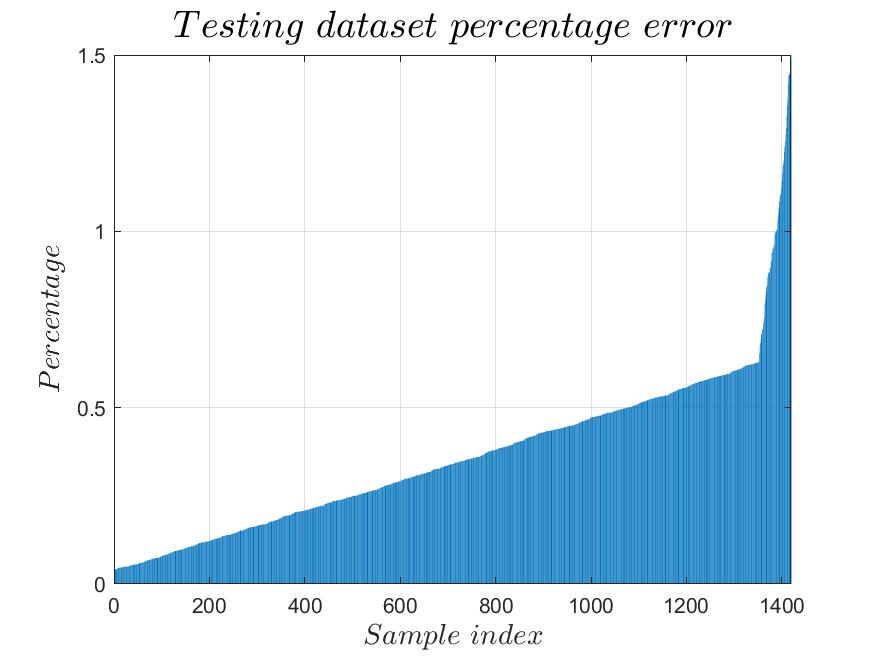}
        \caption{}
        \label{fig:sub15}
    \end{subfigure}
    \hfill
    \begin{subfigure}[b]{0.241\textwidth}
        \centering
        \includegraphics[width=\linewidth]{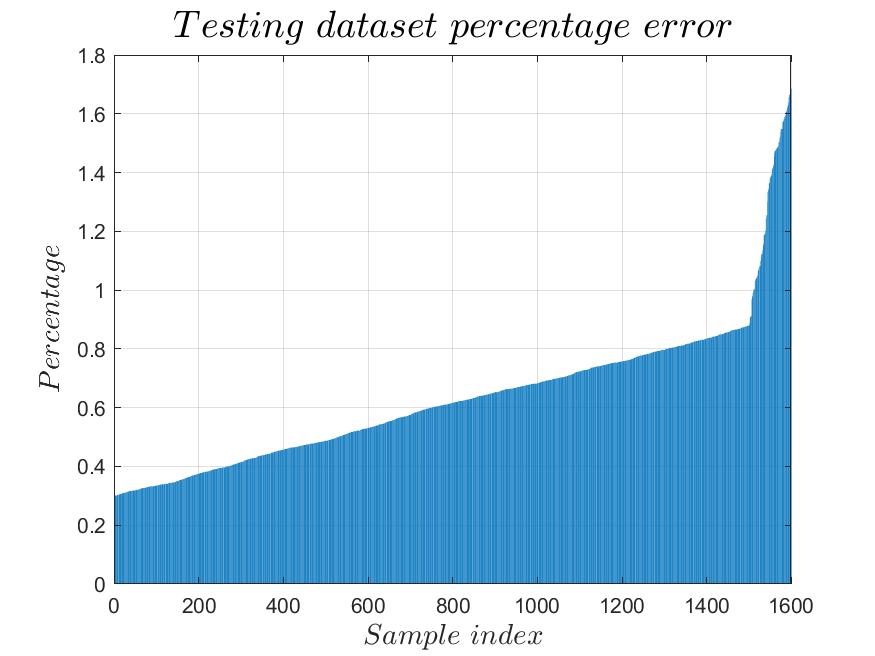}
        \caption{}
        \label{fig:sub25}
    \end{subfigure}
    \caption{Testing dataset percentage error between unit commitment costs obtained by traditional and SVM surrogate TSUC models: (a) 57-bus system, (b) 118-bus system.}
    \label{barchart}
\end{figure}

The percentage error between the actual TSUC cost and the SVM-based surrogate model cost for the testing datasets is illustrated in Fig.~\ref{barchart}. The data points are sorted in ascending order of error magnitude. For the 57-bus system, most samples exhibit errors well below 1.0\%, with only a few outliers approaching the upper bound of 1.5\%. The average error is 0.63\%. Similarly, for the 118-bus system, most samples remain under 1.0\% error, although the maximum deviation reaches 1.7\%. The average error is 0.88\%.

\vspace{-6pt}

\section{Discussion}
\label{s4}
\subsection{Performance Analysis of the SVM Surrogate Model}
The results presented in this study demonstrate the exceptional capability of linear SVM classifiers to serve as effective surrogate models for network constraint feasibility assessment in TSUC problems. The achieved classification accuracies indicate that the linear kernel successfully captures the underlying feasibility boundaries in the DCOPF solution space. More detailed analysis of the confusion matrix reveals promising performance characteristics.
The extremely low false positive rates are critical from a power system security perspective. False positives in this context represent scenarios where the surrogate model incorrectly identifies an infeasible operating point as feasible, potentially leading to constraint violations if implemented in real-time operations. The observed false positive rates of less than 0.5\% provide a high degree of confidence in the surrogate model's reliability for maintaining system security constraints.

\subsection{Computational Efficiency and Scalability}
The computational time reductions of 46\% and 31\% for the 57-bus and 118-bus systems, respectively, highlight the value of the proposed surrogate-assisted optimization framework. The consistent computational savings across different system sizes demonstrate the robustness and reliability of the approach. Even as system complexity increases, the framework maintains significant efficiency gains, with the 118-bus system still achieving a considerable 31\% reduction in computational time.
This sustained performance across varying scales indicates strong potential for real-world power system applications. The observed scaling characteristics provide valuable insights for implementation strategies: while smaller systems benefit from the highest relative savings, larger systems still achieve substantial absolute time reductions that translate to meaningful operational improvements. For very large-scale systems (e.g., thousands of buses), these findings suggest promising opportunities for enhanced implementation through complementary strategies such as hierarchical decomposition or hybrid approaches, which could amplify the already significant computational benefits demonstrated by the surrogate-assisted framework.

\subsection{Accuracy of SVM-Constrained TSUC Generation Cost}
The thermal unit generation costs, as illustrated in Figs. \ref{a} and \ref{b}, demonstrate high accuracy across both test systems. The close alignment between the SVM surrogate model generation cost (blue triangles) and the traditional TSUC with line flow limits (red circles) across all time periods validates the effectiveness of the linear SVM approach. The maximum absolute errors observed remain significantly low, corresponding to relative percentage errors well below 1\%, which are comfortably within acceptable bounds for practical power system operations.
The temporal consistency of the SVM-constrained TSUC generation costs, showing similar patterns and magnitudes across the 24-hour operational horizon, indicates that the surrogate model maintains its reliability across varying load conditions and operational scenarios. This sustained performance throughout different operational states further reinforces the robustness of the surrogate-assisted framework for real-world scheduling applications.

\subsection{Implications for Power System Operations}
Integrating SVM-based surrogate models into TSUC optimization presents several practical advantages for power system operators. First, the significant reduction in computational time enables more frequent re-optimization, allowing for better adaptation to changing system conditions and improved incorporation of renewable energy forecasts. Second, the high accuracy of the surrogate model ensures that operational decisions maintain the required level of system security while achieving near-optimal economic dispatch.
The asymmetric nature of the misclassification errors, with slightly higher false positive rates than false negative rates, reflects the inherent conservatism built into the training process. This characteristic benefits power system applications, as it requires caution when dealing with system security constraints.

\subsection{Limitations and Future Research Directions}
While the results are highly encouraging, several limitations warrant discussion. The study focuses on DCOPF formulations, which neglect reactive power flows and voltage constraints. Future research should investigate the applicability of the SVM surrogate approach to ACOPF problems, which would provide a more comprehensive representation of power system physics.
The impact of renewable energy integration, with its associated variability and uncertainty, on the surrogate model performance remains to be thoroughly investigated.
The observed trend of diminishing computational returns with increased system size suggests hybrid approaches combining SVM surrogates with other acceleration techniques (such as network reduction or decomposition methods) may be necessary for large-scale systems. Furthermore, the dynamic updating of surrogate models to adapt to changing system conditions and topology modifications represents an important area for future development.

\subsection{Practical Implementation Considerations}
Successfully deploying SVM-based surrogate models in operational environments requires careful consideration of several factors. The training data quality and coverage are crucial for robust performance across diverse operating conditions. The approach would benefit from continuous learning frameworks that update the surrogate model as new operational data becomes available.
While not explicitly quantified in this study, the computational overhead of the initial training phase represents a one-time investment that pays dividends through repeated use in operational optimization. The linear nature of the SVM kernel simplifies the hyperplane extraction process, making the integration into existing optimization frameworks relatively straightforward.

\section{Conclusion}
\label{s5}
This paper presented a machine learning surrogate modeling framework for mitigating the computational challenges of two-stage stochastic unit commitment under renewable energy uncertainty. By leveraging the polyhedral structure of DC power flow constraints, a learning-based surrogate was constructed with the governing equations of the SVM classifier to represent the network feasibility design space using a single halfspace. This approach significantly reduces the dimensionality of the optimization problem while preserving physical interpretability and mathematical tractability.

The proposed framework was justified theoretically and empirically. It was demonstrated that, under the DC approximation, line flow constraints can be decoupled from binary commitment decisions, enabling the use of SVM classifiers trained on DCOPF data. Case studies on the IEEE 57- and 118-bus systems validated the surrogate model accuracy—exceeding 99.7\%—and confirmed its ability to maintain reliability with minimal risk of constraint violations. Furthermore, integrating the surrogate into TSUC results in computational speedups (46\% and 31\% for the two systems, respectively) while retaining negligible generation cost errors.

The results affirm the potential of interpretable machine learning models in enhancing the scalability of stochastic power system optimization without compromising security or economic performance. Future work may extend this framework to ACOPF formulations, incorporate dynamic model updates to account for evolving grid conditions, and explore hybrid surrogate strategies for large-scale systems. Overall, this study contributes a computationally efficient and theoretically grounded approach that supports real-time decision-making in modern, uncertainty-aware power systems.

\bibliographystyle{IEEEtran}
\bibliography{references}

\end{document}